\documentclass[12pt]{iopart}
\usepackage[dvips]{graphicx}
\usepackage{cite}
\begin{document}
\title[Sawtooth control using ECCD in ASDEX Upgrade]{Sawtooth control using electron cyclotron current drive in the presence of energetic particles in high performance ASDEX Upgrade plasmas}

\author{IT Chapman$^{1}$, V Igochine$^{2}$, M Maraschek$^{2}$, PJ McCarthy$^{3}$, G Tardini$^{2}$, the ASDEX Upgrade ECRH Group and the ASDEX Upgrade Team}

\address{$^{1}$ Euratom/CCFE Fusion Association, Culham Science Centre, Abingdon, OX14 3DB, UK \\
$^{2}$  Max-Planck Institut f\"{u}r Plasmaphysik, Euratom-Association, D-85748 Garching, Germany \\
$^{3}$ Department of Physics, University College Cork, EURATOM-Association DCU, Cork, Ireland}

\ead{ian.chapman@ccfe.ac.uk}

\begin{abstract}

Sawtooth control using steerable electron cyclotron current drive (ECCD) has been demonstrated in ASDEX Upgrade plasmas with a significant population of energetic ions in the plasma core and long uncontrolled sawtooth periods.
The sawtooth period is found to be minimised when the ECCD resonance is swept to just inside the $q=1$ surface.
By utilising ECCD inside $q=1$ for sawtooth control, it is possible to avoid the triggering of neoclassical tearing modes, even at significnatly higher pressure than anticipated in the ITER baseline scenario.
Operation at 25\% higher normalised pressure has been achieved when only modest ECCD power is used for sawtooth control compared to identical discharges without sawtooth control when neo-classical tearing modes are triggered by the sawteeth.
Modelling suggests that the destabilisation arising from the change in the local magnetic shear caused by the ECCD is able to compete with the stabilising influence of the energetic particles inside the $q=1$ surface.

\end{abstract}
\maketitle
%---------------------------------------------------------------------------
%---------------------------------------------------------------------------
\section{Introduction}

Sawtooth oscillations in tokamak plasmas are characterised by quasi-periodic collapses in the temperature and density in the plasma core \cite{vonGoeler}.
The drop in fusion performance caused by sawteeth is not of significant concern for ITER; however, the triggering of more deleterious instabilities, such as neoclassical tearing modes, means that sawtooth control remains an important issue.
A typical sawtooth cycle exhibits three phases: (i) the sawtooth ramp phase during which the plasma density and temperature increase approximately linearly with respect to time; (ii) the precursor phase, during which a helical magnetic perturbation grows until (iii) the fast collapse phase, when the density and temperature drop rapidly.
In order to control sawteeth one must use actuators which can affect the second phase in the cycle - the trigger of the instability growth.
The sawtooth is thought to be caused by the growth of a $n=m=1$ internal kink mode -- a fundamental magnetohydrodynamic (MHD) oscillation of the form $\xi \sim \exp(im\theta - in\phi)$ where $m$ and $n$ are the poloidal and toroidal mode number respectively, $\xi$ is the perturbation to the plasma and $\theta$ and $\phi$ are the poloidal and toroidal angles.

Minority populations of super thermal ions are predicted analytically and demonstrated experimentally to delay the onset of the $n/m=1/1$ internal kink mode, thereby increasing the period between sawtooth crashes.
The presence of fusion-born alpha particles in ITER is predicted to significantly lengthen the sawtooth period \cite{Hu,Porcelli1996,ChapmanEPS,Chapman2011}, which has been shown empirically to result in an increased likelihood of triggering NTMs \cite{Chapman2010,Buttery2004}. 
Consequently a control scheme which can maintain small, frequent sawtooth crashes which avoid seeding deleterious NTMs whilst still flushing irradiating impurities from the plasma core is necessary in ITER.

The fundamental trigger of the sawtooth crash is thought to be the onset of an $m=n=1$ mode, although the dynamics of this instability are constrained by many factors including not only the macroscopic drive from ideal MHD, but collisionless kinetic effects related to high energy particles \cite{Porcelli1991,Graves2009,Graves2003} and thermal particles \cite{Kruskal,Fogaccia}, as well as non-ideal effects localised in the narrow layer around $q=1$.
Here, the safety factor is $q=\textrm{d}\psi_{\phi}/\textrm{d}\psi_{\theta}$ and the magnetic shear is $s=(r/q) \textrm{d}q/\textrm{d}r$ with $\psi_{\theta}$ and $\psi_{\phi}$ the poloidal and toroidal magnetic fluxes respectively.
A heuristic model predicts that a sawtooth crash will occur in the presence of energetic ions when various criteria are met \cite{Porcelli1996,SauterVarenna,Cowley}, with the defining one usually given in terms of a critical magnetic shear determined either by the pressure gradient, $s_{1}>s_{crit}(\omega_{*i})$, or by the mode potential energy, written as:
\begin{equation}
s_{1} > \textrm{max} \Big(\frac{4\delta W}{\xi_{0}^{2} \epsilon_{1}^{2} R B^{2} c_{\rho} \hat{\rho}}, s_{crit}(\omega_{*i}) \Big)
\label{eq:Porcelli4}
\end{equation}
where $c_{\rho}$ is a normalisation coefficient of the order of unity, $\hat{\rho}=\rho_{i}/r_{1}$, $\rho_{i}$ is the ion Larmor radius, $R$ is the major radius, $B$ is the toroidal field, $\epsilon_{1}=r_{1}/R$, $r_{1}$ is the radial position of the $q=1$ surface, $\omega_{*i}$ is the ion diamagnetic frequency and $\xi_{0}$ is the magnetic perturbation at the magnetic axis.
The change in the kink mode potential energy is defined such that $\delta W = \delta W_{core} + \delta W_{h}$ and $\delta W_{core}=\delta W_{f} + \delta W_{KO}$ where $\delta W_{KO}$ is the change in the mode energy due to the collisionless thermal ions \cite{Kruskal}, $\delta W_{h}$ is the change in energy due to the fast ions and $\delta W_{f}$ is the ideal fluid mode drive \cite{Bussac}. 
It is clear that, if this model for sawtooth onset is correct, the sawteeth can be deliberately stimulated by increasing the local magnetic shear at $q=1$, $s_{1}$.

When electron cyclotron resonance heating (ECRH) is applied to the plasma, the local current density changes because the temperature also changes, and subsequently causes a change in the conductivity. 
Adding a toroidal component to the wave vector of the launched EC waves results in an electron cyclotron driven current either parallel (co-ECCD) or anti-parallel (counter-ECCD) to the Ohmic current.
When applied near the $q=1$ surface, the radius of the $q=1$ surface, $r_{1}$, is moved and the magnetic shear at $q=1$, $s_{1}$, is changed, thus affecting the likelihood of a sawtooth crash according to equation \ref{eq:Porcelli4}. 

Control of the sawtooth period with ECCD has been demonstrated on a number of tokamaks \cite{Ahkaev,Muck,Angioni,Ikeda,Pinsker,Westerhof,Lennholm2007}, and consequently has been included in the design of the sawtooth control system for ITER \cite{ITERECH,Darbos}.
The history of sawtooth control using current drive is reviewed in \cite{ChapmanRev}.
The suppression of sawteeth for NTM prevention using ECCD has been demonstrated directly at high pressure on ASDEX Upgrade by using co-ECCD just outside the $q=1$ surface \cite{Maraschek}.
At the end of the gyrotron pulse, a sawtooth crash occurred and an NTM was triggered, resulting in substantial degradation of the plasma performance.
That said, it is widely accepted that sawteeth are unlikely to be avoided throughout an ITER discharge, and so a similar demonstration of avoidance of NTMs in high performance plasmas with deliberately-paced frequent sawteeth is required.
An additional benefit of using ECCD for sawtooth control to avoid NTMs is that the ECH is directed well inside $q=1$ and so is usefully heating the core of the plasma.
Conversely if the ECCD is used to suppress NTMs at higher rational surfaces, notably at $q=2$, the power is not used for heating and so significantly reduces the fusion yield, $Q$ \cite{Sauter2010}.

The remaining concern about sawtooth control achieved by current drive is whether changes in $s_{1}$ can overcome the stabilisation due to energetic particles.
In ITER, there is likely to be a large stabilising potential energy contribution, $\delta W_{h}$ in equation \ref{eq:Porcelli4} due to the fusion-born $\alpha$ particles.
Combining this with the small $\hat{\rho}$ in the denominator means that the critical shear at which the internal kink mode becomes unstable is increased.
Consequently, recent experiments have focussed on destabilising sawteeth when there is a significant population of energetic particles in the plasma core.
Sawtooth destabilisation of long period sawteeth caused by fast ions with energies $\geq 0.5$MeV arising from ICRH was achieved in Tore Supra, even with a modest level of ECCD power  \cite{Lennholm,Lennholm2009}.
Similarly, ECCD destabilisation has also been achieved despite ICRH-accelerated neutral beam injection (NBI) ions in the core of ASDEX Upgrade \cite{Igochine2010} as well as with normal NBI fast ions in  DIII-D \cite{Chapman2012}, ASDEX Upgrade \cite{Muck} and JT-60U \cite{Isayama}.
Despite these promising results, demonstration of NTM avoidance through sawtooth control in the presence of energetic particles with steerable ECCD has yet to be demonstrated in ITER-like conditions; this is what is addressed in this paper.
In section \ref{sec:control} sawtooth control in high performance ASDEX Upgrade plasmas is shown in the presence of energetic NBI ions and the optimal resonance position to minimise the sawtooth period is found by sweeping the EC launching mirrors.
After demonstrating the optimal deposition for ECCD in order to destabilise the sawteeth, the improvement in fusion performance with active sawtooth control is discussed in section \ref{sec:highbeta}.
In section \ref{sec:modelling} the effect of changing the magnetic shear is compared to the stabilising drive from the fast ions using numerical simulation, before the implications of this work are discussed in section \ref{sec:discussion}.

%---------------------------------------------------------------------------
\section{Sawtooth Control using ECCD in the presence of energetic ions} \label{sec:control}

\begin{figure}
\begin{center}
\includegraphics[width=0.9\textwidth]{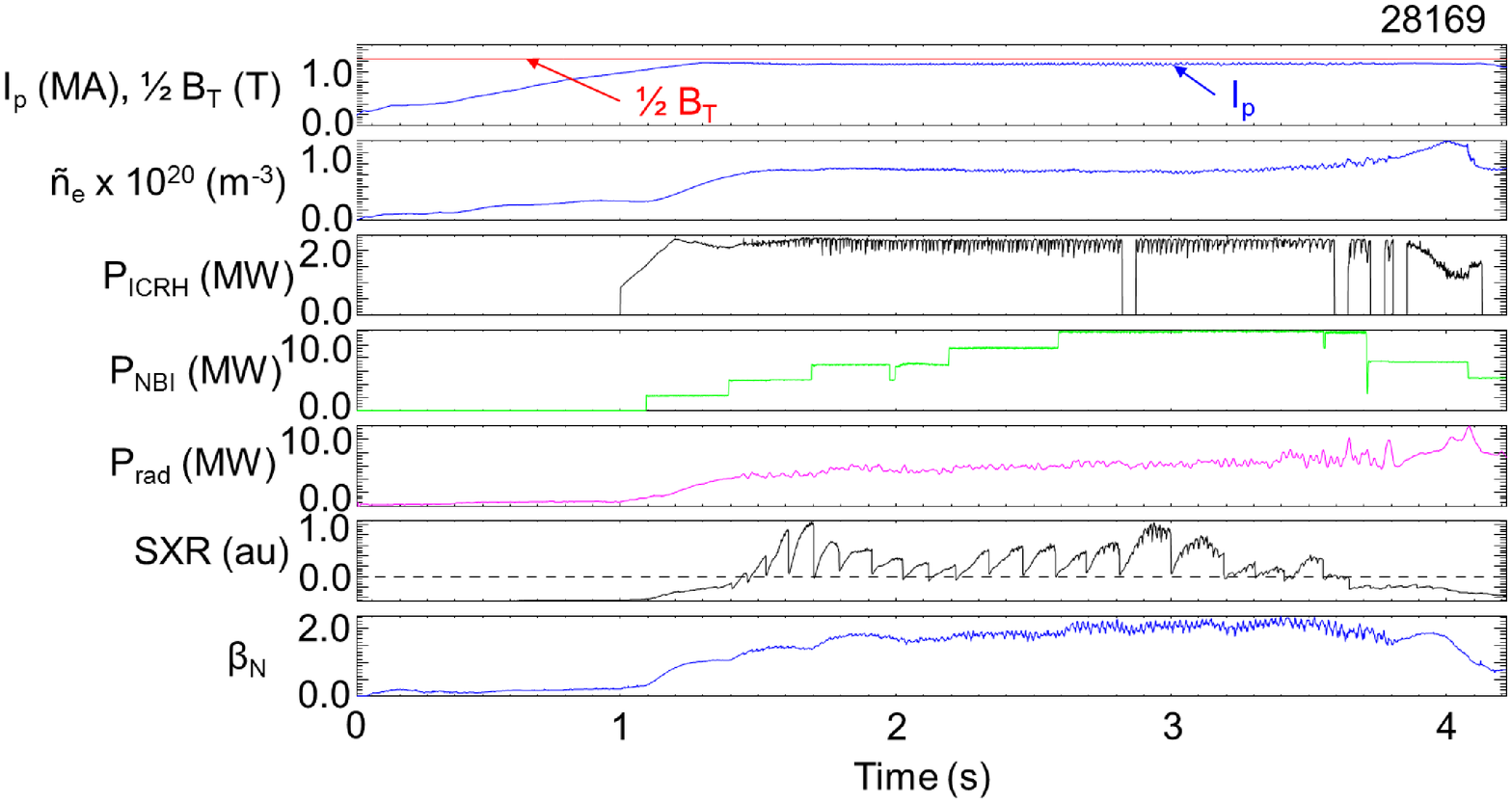}
\end{center}
\caption{Time traces for a typical high performance ASDEX Upgrade plasma -- discharge 28169 -- showing: The plasma current and half of the toroidal field; the line averaged density which is relatively constant over the sawtoothing period; the ICRH heating power; the NBI heating power; the radiated power; the Soft X-ray emission from a central channel; and the normalised plasma pressure, $\beta_{N}$ which fractionally exceeds the ITER target value of 1.8.}
\label{fig:28169}
\end{figure}

In order to replicate typical ITER operational conditions with a significant population of energetic ions in the plasma core, ASDEX Upgrade \cite{Herrmann} plasmas can be heated with neutral beam injection (NBI) and ion cyclotron resonance heating (ICRH).
Figure \ref{fig:28169} shows a typical ASDEX Upgrade discharge with 2MW of ICRH combined with 12MW of NBI operating at $\beta_{N} \sim 2$ just above the target normalised pressure for ITER baseline plasmas (where $\beta_{N} \sim 1.8$) and $H_{98,y2}=0.9$.
Here $H_{98,y2}$ is the energy confinement enhancement factor, $\beta_{N}= \beta a B_{0}/I_{p}$ where $a$ is the minor radius, $I_{p}[\textrm{MA}]$ is the plasma current, $\beta = 2\mu_{0} \langle p \rangle / B_{0}^{2}$ and $\langle \cdots \rangle$ represents a volume average and $p$ is the plasma pressure.
The magnetic field ($B_{T}=2.5$T) and current ($I_{p}=1.1$MA) mean that the safety factor at the 95\% flux surface is $q_{95}=3.9$. This is above the ITER design value of 3.0, but this was necessary in order to have the ICRH resonance position in the plasma core and the ECRH resonance off-axis.
Nonetheless, the $q$-profile has a broad low-shear region with the radial position of the $q=1$ surface is $\rho_{1}>0.3$, which is approaching the value of $\rho_{1}=0.45$ expected in ITER.
The plasma illustrated in figure \ref{fig:28169} experiences long period sawteeth throughout $\tau_{saw} \approx 150$ms compared with an energy confinement time of $\tau_{E} \approx 80$ms.
Scaling the sawtooth period by the resistive diffusion time \cite{Park} and $r_{1}$, this period is roughly equivalent to 45s in ITER, which is approaching the expected critical sawtooth period likely to seed NTMs \cite{Chapman2010}.
It should be noted that with 12MW of uni-directional neutral beam heating, there is significant NBI-induced torque leading to a much faster toroidal rotation frequency than anticipated in ITER.
This differential rotation is likely to inhibit the triggering of NTMs by 1/1 internal kinks.
Finally, an important difference between these plasmas and the ITER baseline scenario is the fraction of fast ions: The NBI and ICRH induced fast ions in these ASDEX Upgrade plasmas constitute approximately 20\% of the stored energy, whilst the fusion-born alphas and heating-induced fast ions in ITER result in a fast ion fraction, $\langle \beta_{h}\rangle /\langle \beta \rangle$, approaching 45\% ($\beta_{\alpha}$ from \cite{Budny2002}, $\beta_{NBI}$ from \cite{Asunta}). 

The plasma performance in discharge 28169 shown in figure \ref{fig:28169} is limited by the appearance of an $m/n=2/1$ tearing mode, triggered by a long sawtooth period at 2.3s.
Whilst this NTM is not disruptive, it means that the increments in NBI power at both 2.2s and 2.5s lead to little improvement in the normalised pressure as the NTM progressively degrades the confinement, giving rise to the confinement enhancement factor of $H_{98,y2}=0.9$ below the ITER baseline assumption of $H_{98,y2}=1.0$.
It is exactly this situation -- the lengthening of the sawtooth period by the presence of fast particles, leading to the triggering of NTMs which persistently degrade performance -- which sawtooth control with ECCD aims to avoid, permitting higher performance and confinement improvement.

\begin{figure}
\begin{center}
\includegraphics[width=0.7\textwidth]{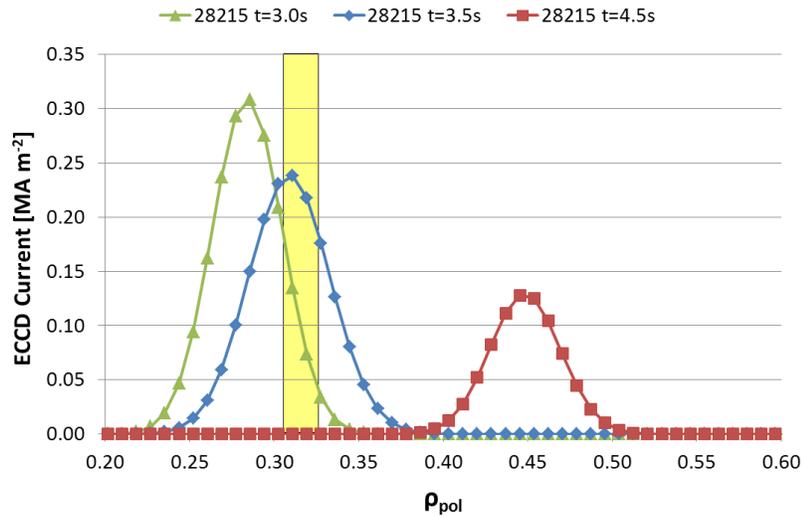}
\end{center}
\caption{The ECCD driven current profile for ASDEX Upgrade discharge 28215 predicted by the \textsc{Torbeam} code \cite{Poli} when the resonance is inside, around and well outside the inversion radius found from the Soft X-ray emission, marked by the shaded region.}
\label{fig:ECCD}
\end{figure}

\begin{figure}
\begin{center}
\includegraphics[width=0.7\textwidth]{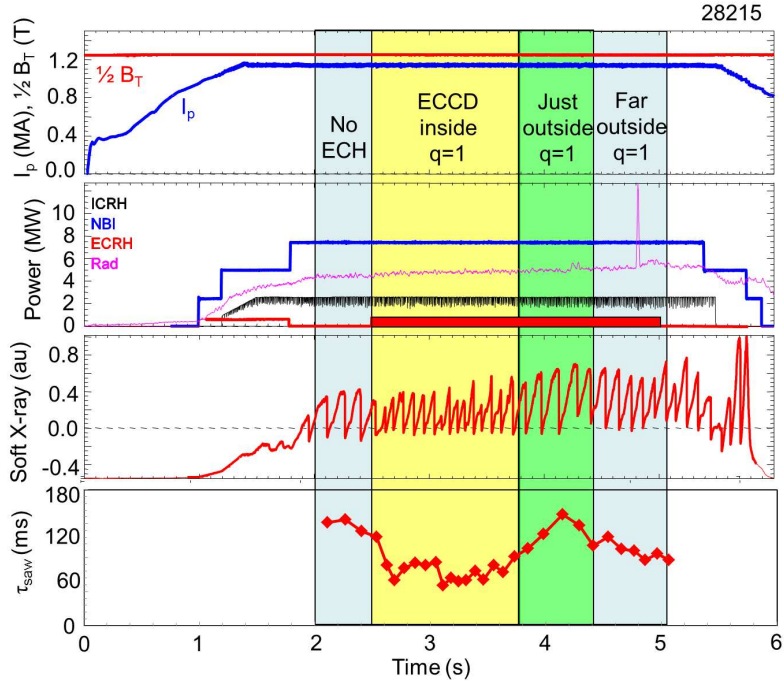}
\end{center}
\caption{Time traces for ASDEX Upgrade discharge 28215 as the ECCD is swept from inside to outside the $q=1$ surface, showing: the plasma current and half the toroidal field; the ICRH, NBI, ECRH and radiated power; the soft X-ray emission from a central channel; and the sawtooth period as a function of time. The coloured bands indicate the position of the EC resonance predicted by the \textsc{Torbeam} code with respect to the inversion radius found from the Soft X-ray emission. The sawtooth period is minimised when the ECCD is inside $q=1$.}
\label{fig:28215}
\end{figure}

In order to find the optimal position for driven current to change the local magnetic shear and so destabilise the sawteeth, a sweep of the ECCD was performed using the steerable EC mirrors.
A sweep was performed moving the EC resonance from $\rho_{dep}=0.2$ to $\rho_{dep}=0.45$ in 2.5s to cross the $q=1$ surface, which was at $\rho_1 \sim 0.32$.
Figure \ref{fig:ECCD} shows the EC driven current profile predicted by the \textsc{Torbeam} code \cite{Poli} when the resonance is inside, around and well outside the inversion radius found from the Soft X-ray emission.
In all cases, the toroidal component to the wave vector is such that the ECCD is reasonably narrow and affects only a small region of the current density profile. 
Previously it was shown that ECCD is more effective for sawtooth control than ECH in ASDEX Upgrade \cite{Muck}, so we concentrate on the effects of ECCD in this paper.
The sawtooth behaviour during this ECCD sweep is shown in figure \ref{fig:28215}.
The plasma is heated with 2MW of core ICRH and only 7MW of NBI so that the pressure is lower than in typical high-performance plasmas (like in figure \ref{fig:28169}). 
This facilitates a full sweep of the EC deposition to outside $q=1$, which would almost certainly incur triggering of NTMs at higher pressure as it leads to longer sawtooth periods.
The sawtooth period is clearly decreased when the ECCD is inside $q=1$, with the minimum in the sawtooth period occurring when the EC deposition is inside $q=1$, as expected \cite{Angioni,ChapmanRev}.
As the resonance is swept across $\rho_{1}$ the sawtooth period lengthens above the level before the ECCD was applied, before returning to approximately the pre-ECCD level once more when the driven current is well outside $\rho_{1}$ and no longer affecting the local magnetic shear, $s_{1}$.
The fact that the sawtooth period when the ECCD is inside $q=1$ is consistently around half the level of that before the ECCD is applied, irrespective of the exact deposition level, means that robust control is likely to be achievable without requiring fine deposition feedback control.
This relative insensitivity to the precise EC resonance position is utilised in the next section to optimise performance by applying ECCD optimised from this study, but at higher plasma pressures where the Shafranov shift, and correspondingly the $q=1$ radius, are different.
The fact that a very marked destabilisation of the sawteeth is observed in the presence of a relatively significant fraction of fast ions, with some very energetic particles born due to RF heating, is encouraging for the applicability of ECCD sawtooth control in ITER.

%---------------------------------------------------------------------------
\section{Optimising performance using ECCD sawtooth control in ITER demonstration plasmas} \label{sec:highbeta}

\begin{figure}
\begin{center}
\includegraphics[width=0.7\textwidth]{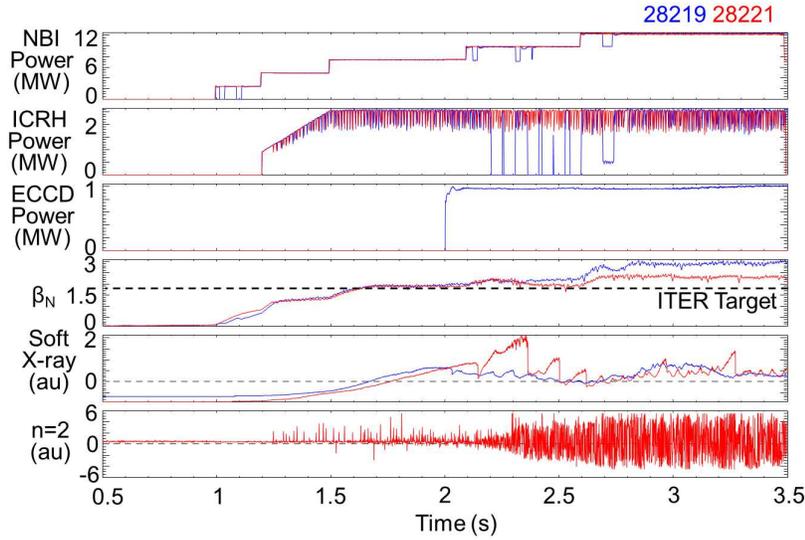}
\end{center}
\caption{A comparison of ASDEX Upgrade discharges 28219 which has ECCD applied for sawtooth control and 28221 which does not. The time traces show the NBI heating power; the ICRH heating power; the ECCD power with the resonance position held fixed inside $q=1$; the normalised beta compared to the ITER target value; the soft X-ray emission from a central channel; and the $n=2$ mode activity measured by magnetic pick-up sensors. In shot 28221 a 3/2 NTM is triggered by a long sawtooth at 2.15s and results in much lower plasma performance than shot 28219 where the sawtooth period is very short throughout.}
\label{fig:28219_28221}
\end{figure}

The sweeps of the EC deposition with the steerable mirrors outlined in section \ref{sec:control} allowed the optimal EC resonance location to be inferred. 
The relative insensitivity of the minimum in sawtooth period to the exact resonance location with respect to $q=1$ means that sawtooth control in ASDEX Upgrade can be achieved without real-time steering of the EC launcher mirrors.
Indeed, this insensitivity allows the resonance location to be fixed and the pressure to be increased, whilst retaining frequent, controlled sawteeth despite the enhanced Shafranov shift at higher $\beta$. 
The EC mirror settings were fixed such that $\rho_{1}-\rho_{dep} \approx 0.1$, then the auxiliary NBI power was incremented step-wise to increase the pressure and examine the efficacy of the sawtooth control for avoiding NTMs.
Figure \ref{fig:28219_28221} shows two identical ASDEX Upgrade plasmas, one with 1MW of core ECCD to keep the sawtooth period small (shot 28219), and one without core ECCD (shot 28221).
In the absence of sawtooth control, a 2/1 NTM is triggered by a sawtooth crash at t=2.15s at $\beta_{N}=2$, only 10\% above the ITER operating normalised pressure.
Conversely, in shot 28219, only 1MW of ECCD inside $q=1$ drives the internal kink mode unstable resulting in small, frequent sawteeth and avoiding NTMs throughout the discharge, even as the normalised pressure is increased to $\beta_{N}=3.0$.
The energy confinement enhancement factor reaches $H_{98,y2}^{28219}=1.25$ when the sawtooth control is applied, compared to only $H_{98,y2}^{28221}=0.95$ without ECCD mode control.
Just as in DIII-D \cite{Chapman2012}, a low level of core ECCD allows a sawtoothing plasma with a broad low-shear $q$-profile to reach much higher normalised pressure than anticipated in ITER without exhibiting tearing modes, although in these ASDEX Upgrade plasmas the fast ion fraction and maximum energy are even higher than in previous results.

\begin{figure}
\begin{center}
\includegraphics[width=0.7\textwidth]{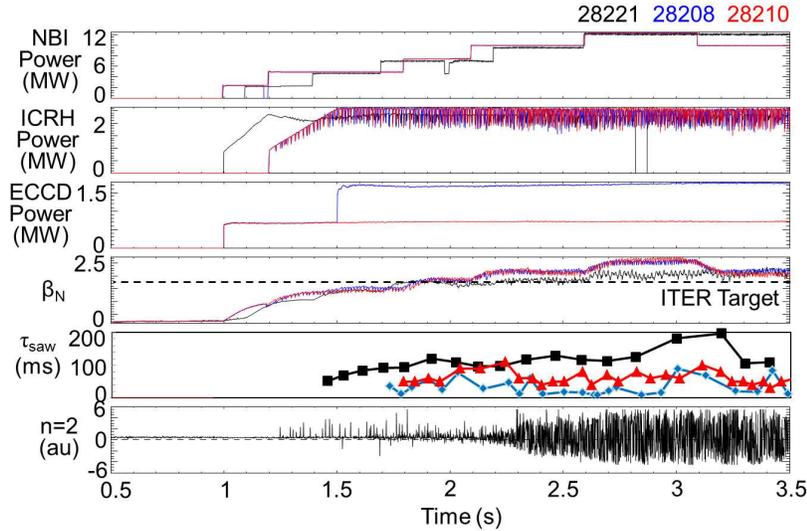}
\end{center}
\caption{A comparison of the efficacy of ECCD sawtooth control at different ECCD power levels. The time traces show the NBI heating power; the ICRH heating power; the ECCD power used for sawtooth control with the resonance position fixed inside $q=1$; the normalised beta compared to the ITER target value; the sawtooth period found from the soft X-ray emission and the $n=1$ mode activity for shot 28221 where a 3/2 tearing mode is triggered by a sawtooth crash. The sawtooth period and achievable pressure are similar in discharges 28208 and 28210 despite different ECCD power levels, indicating that only a small driven current is required for efficient NTM avoidance.}
\label{fig:28208_28210}
\end{figure}

Figure \ref{fig:28208_28210} shows three very similar ASDEX Upgrade discharges with different core ECCD power levels: 28169 which has no core ECH, 28210 which has 0.8MW of ECCD inside $\rho_{1}$ to control sawteeth and 28208 which has 1.7MW of ECCD.
In the absence of ECCD for sawtooth control, a 2/1 NTM is triggered by a long sawtooth period at $t=2.32$s, limiting the $\beta_{N}$ achieved thereafter.
In contrast, both plasmas with ECCD achieve short sawtooth periods throughout, consistently less than 100ms, and as a result avoid NTMs allowing 25\% higher normalised pressure to be attained.
Whilst the higher ECCD power in shot 28208 does result in the lowest sawtooth period, it does not achieve a higher pressure since the sawteeth themselves do not deleteriously affect performance and one only needs to avoid the triggering of NTMs, which shot 28210 also achieves despite lower ECCD power. 
It is evident from figures \ref{fig:28219_28221} and \ref{fig:28208_28210} that sawtooth control with ECCD is effective for the avoidance of NTMs at auxiliary heating levels which would otherwise incur a sawtooth-triggered NTM. 
The NBI power is increased progressively in both figures to the maximum available power in these ASDEX Upgrade plasmas and NTMs are avoided at all levels of heating (and thus fast ion fraction).

%---------------------------------------------------------------------------
\section{Modelling the effect of ECCD in high performance, high fast ion fraction plasmas} \label{sec:modelling}

In order to assess whether the change in the local magnetic shear produced by electron cyclotron current drive is responsible for the sawtooth control reported in sections \ref{sec:control} and \ref{sec:highbeta}, the linear stability of the internal kink mode has been assessed.
Although such linear analysis cannot be used to infer the behaviour of the nonlinear sawtooth period, it is indicative of the sawtooth stability, and has been used to make successful experimental comparisons of sawtooth behaviour in MAST \cite{Chapman2006}, TEXTOR \cite{ChapmanTEXTOR}, JET \cite{Chapman2007,Chapman2008}, DIII-D \cite{Chapman2012} ASDEX Upgrade \cite{Chapman2009,ChapmanPoP}.

\begin{figure*}
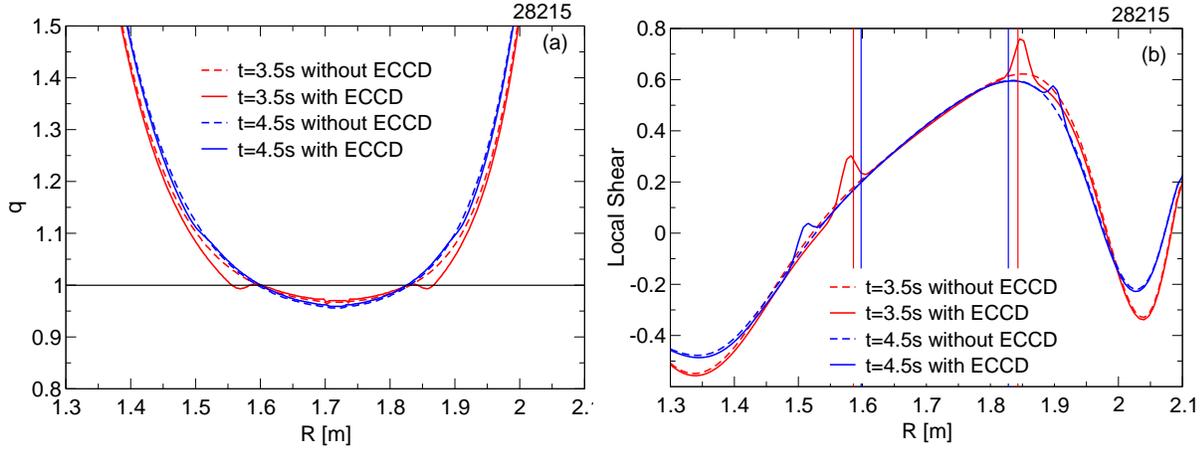

\hspace{-0.2cm}
\vspace{0.5cm}
\begin{minipage}{0.5\textwidth}
\begin{center}
\includegraphics[width=\textwidth]{figure6a.eps}
\end{center}
\end{minipage}
\begin{minipage}{0.5\textwidth} % A minipage that covers half the page
\begin{center}
\includegraphics[width=\textwidth]{figure6b.eps}
\end{center}
\end{minipage}
\caption{The (a) $q$-profile and (b) local magnetic shear as a function of major radius as calculated by \textsc{Cliste} with (solid lines) and without (dashed lines) including the ECCD profile predicted by \textsc{Torbeam} in the current density for ASDEX Upgrade shot 28215 at t=3.5s (ECCD at $q=1$ giving optimal sawtooth destabilisation) and t=4.5s (ECCD well outside $q=1$). The vertical lines on the local shear profile show the positions of the $q=1$ surface. The ECCD only affects the current density, $q$, and shear profiles in a narrow local region at the deposition location.}
\label{fig:qs}
\end{figure*}

Fixed-boundary equilibria are reconstructed using the \textsc{Helena} code \cite{Huysmans}, taking as input the current density profile from the \textsc{Cliste} equilibrium code \cite{McCarthy}, which itself is constrained to include the ECCD profiles predicted by \textsc{Torbeam} and the $q=1$ surface position inferred from the inversion radius found from soft X-ray emission.
The ECCD was incorporated in \textsc{Cliste} reconstruction by adding the ECCD current density profile from \textsc{Torbeam} to the $f \textrm{d}f/\textrm{d}\psi (\psi)$ source function which forms part of the interpretive run output. 
Subsequently predictive \textsc{Cliste} runs were made with and without the ECCD-modified $f \textrm{d}f/\textrm{d}\psi$ profile.  
The resultant ECCD current density bump can then be back-compared to the \textsc{Torbeam} prediction with good agreement -- for instance at t=3.5s the ECCD feature has a peak value of 0.26 MA/$\textrm{m}^2$, which is just 5\% difference to the peak value of 0.245 MA/$\textrm{m}^2$ from \textsc{Torbeam}, as seen in figure \ref{fig:ECCD}.
This technique produces the safety factor and local magnetic shear profiles illustrated in figure \ref{fig:qs}.
Here the local shear is defined as \cite{Nadeem}
\begin{equation}
s=-\vec{e}_{\perp} \cdot \nabla \times \vec{e}_{\perp} 
\end{equation}
where $\vec{e}_{\perp} = \nabla \psi / |\nabla \psi | \times \vec{B}/B$ meaning that in cylindrical limit with circular flux surfaces, this approximates to
\begin{equation}
s=\frac{1}{qR} \frac{r}{q} \frac{\textrm{d}q}{\textrm{d}r}
\end{equation}
and can be re-written to be expressed as variables directly calculated by \textsc{Cliste} \cite{McCarthy2}
The increase in the local shear provided by the ECCD is clearly seen in figure \ref{fig:qs}(b).
At $t=$3.5s the ECCD deposition is near the $\rho_{1}$ (found from the inversion radius) and as a result the local shear at $q=1$ (marked by vertical lines in figure \ref{fig:qs}(b)) increases by a factor of two compared to the case without ECCD included in the equilibrium reconstruction. 
Conversely, when the ECCD resonance is outside $q=1$ at $t=4.5$s, the local magnetic shear at $q=1$ is barely affected.

\begin{figure*}
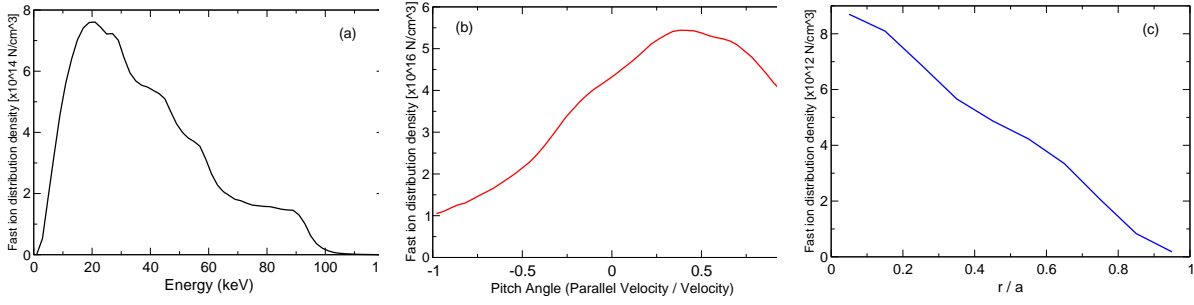

\hspace{-0.2cm}
\vspace{0.5cm}
\begin{minipage}{0.33\textwidth}
\begin{center}
\includegraphics[width=\textwidth]{figure7a.eps}
\end{center}
\end{minipage}
\begin{minipage}{0.33\textwidth} % A minipage that covers half the page
\begin{center}
\includegraphics[width=\textwidth]{figure7b.eps}
\end{center}
\end{minipage}
\begin{minipage}{0.33\textwidth}
\begin{center}
\includegraphics[width=\textwidth]{figure7c.eps}
\end{center}
\end{minipage}
\caption{The fast ion distribution density for ASDEX Upgrade discharge 28219 as calculated by \textsc{Transp} at 3.5s averaged over 20ms as a function of (a) particle energy, (b) pitch angle ($=v_{\parallel}/v$) and (c) radius, where for each plotted variable the fast ion density is integrated over the other two variables.}
\label{fig:fastions}
\end{figure*}

As well as the change in the magnetic shear, the fast ion distribution is also required to assess the change in the potential energy of the internal kink mode.
In order to retain the complex dependence of the fast ion population upon pitch angle, energy and radius, the full Monte Carlo distribution function is employed in the drift kinetic modelling detailed below.
The effect of the fast ions on internal kink stability is analysed using the drift kinetic \textsc{Hagis} code \cite{Pinches}.
\textsc{Hagis} simulates the interaction between the perturbation taken from \textsc{Mishka-F} \cite{ChapmanMISHKA} and the energetic particle distribution taken from the \textsc{Transp} code \cite{Budny1992}.
Figure \ref{fig:fastions} shows the neutral beam fast ion density calculated by \textsc{Transp} as a function of energy, pitch angle and radius when averaged over the other variables in the poloidal plane (ie the energy dependence is integrated across all pitch angles and radii).
The fast ions are peaked near the axis, which is where the NBI is aimed and the ICRH resonance is deposited when $B_{T}=2.5$T.
In the shot considered here, the energetic particle distribution is peaked around $\lambda = v_{\parallel}/v \sim 0.5$ and is approximately Gaussian with respect to their pitch angle at high energies, although at lower energy, the beam ion population tends to isotropy.
The ICRH distribution function is assumed to be bi-Maxwellian in form, as in references \cite{GravesVarenna,Dendy}:
\begin{equation}
f_{h}^{\textrm{ICRH}} = \Big( \frac{m}{2\pi} \Big)^{3/2} \frac{n_{c}(r)}{T_{\perp}(r) T_{\parallel}^{1/2}(r)} \exp \bigg[ -\frac{\mu B_{c}}{T_{\perp}(r)}-\frac{|\mathcal{E}-\mu B_{c}|}{T_{\parallel}(r)} \bigg]
\end{equation}
where the particle energy $\mathcal{E}=mv^{2}/2$, the magnetic moment $\mu=mv_{\perp}^{2}/B$, $\parallel$ and $\perp$ represent the components parallel and perpendicular to the magnetic field respectively, $B_{c}$ is the critical field strength at the resonance and $n_{c}$ is the local density evaluated at $B=B_{c}$.

\begin{figure}
\begin{center}
\includegraphics[width=0.7\textwidth]{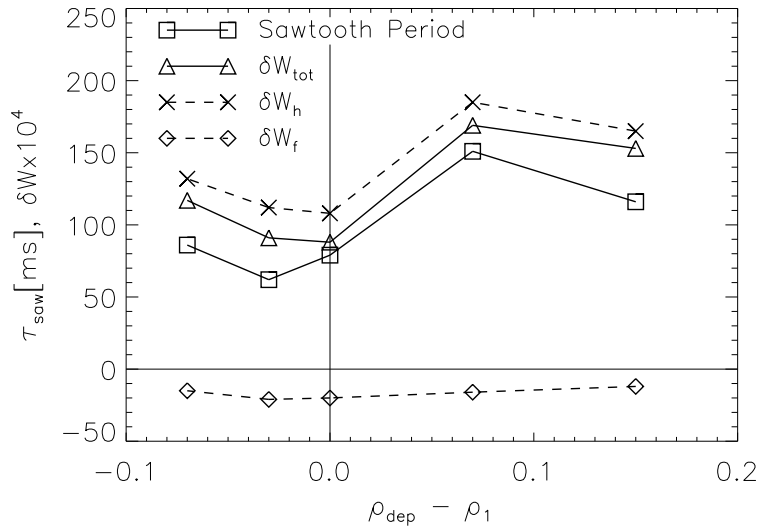}
\end{center}
\caption{The sawtooth period as a function of the peak deposition location of ECCD calculated by \textsc{Torbeam} with respect to the inversion radius in ASDEX Upgrade shot 28215. The sawtooth period dependence is mirrored by the change in the potential energy of the internal kink mode, $\delta W_{tot}$. This is primarily caused by the change in local shear affecting the potential energy arising from fast ions as calculated by \textsc{Hagis}, $\delta W_{h}$.}
\label{fig:dw}
\end{figure}

The effect of changing the local magnetic shear is assessed by calculating the change in the potential energy of the $n=1$ internal kink mode which enters into the critical magnetic shear required for a sawtooth to occur, as given by equation \ref{eq:Porcelli4}. 
The fluid drive for the mode, $\delta W_{f}$ is calculated by \textsc{Mishka-F}, whilst the stabilising effect from the core fast ions, $\delta W_{h}$, resulting from the neutral beam injection is calculated using \textsc{Hagis}.
Figure \ref{fig:dw} shows the sawtooth period for ASDEX Upgrade discharge 28215 (where the ECCD deposition is swept from inside to outside $q=1$ as shown in figure \ref{fig:28215}) as a function of $\rho_{res}-\rho_{1}$.
This is compared to the change in the potential energy of the kink mode as calculated with \textsc{Mishka} and \textsc{Hagis} for the fluid drive and energetic particle response respectively.
When the EC is deposited just inside $\rho_{1}$, the fluid drive for the $n=m=1$ internal kink is maximised because the EC driven current increases both the magnetic shear and $r_{1}$.
As well as driving the internal kink, the stabilising effect of the fast ions is diminished due to the normalisation of $\hat{\delta W}_{h}$ in equation \ref{eq:Porcelli4} by the local magnetic shear.
Here the $\hat{\delta W}_{h}$ is calculated using the fast ion distribution from discharge 28219 throughout, as shown in figure \ref{fig:fastions}.
Whilst linear stability calculations cannot be used to infer the sawtooth period, which is naturally dominated by nonlinear processes, it is indicative of sawtooth stability.
Furthermore, the fact that the change in potential energy of the internal kink, $\delta W_{tot}$, correlates strongly with the sawtooth period gives confidence that the dominant physics is captured in the modelling.
This shows that whilst only a small ECCD power ($\sim$ 1MW) is applied, this provides a significant change in the local magnetic shear near $q=1$ (though a negligible change in the total current), which can counteract the stabilising influence of the population of energetic particles born as a result of 15MW of injected power.

%---------------------------------------------------------------------------
\section{Discussion and conclusions} \label{sec:discussion}

The ELMy H-mode baseline scenario in ITER is expected to experience sawtooth oscillations, and may even require such core reconnection events to alleviate core impurity accumulation.
The fusion born $\alpha$ particles together with fast ions arising from the neutral beam injection and ion cyclotron resonance heating are expected lengthen the sawtooth periods, potentially to the order of 100s \cite{Porcelli1996,Jardin,Onjun}.
Empirical scaling suggests that such long sawteeth are likely to trigger deleterious NTMs \cite{Chapman2010} and therefore active sawtooth control is required.
Whilst NTM suppression is planned for ITER, direct avoidance by sawtooth control is preferable in order to optimise the cost of electricity in a steady-state fusion power plant, since as well as heating the core plasma when ECCD is applied inside $q=1$ for sawtooth control, the ECCD efficiency is far greater near the core than near the $q=2$ surface.
The use of current drive for controlling sawtooth periods is robust and widely demonstrated, but there is little evidence of its use in the presence of significant populations of fast ions which result in a large positive $\delta W_{h}$, which coupled with the small ion Larmor radius, makes the criterion for the necessary magnetic shear challenging (see equation \ref{eq:Porcelli4}).
The results presented here show that not only is ECCD control possible using very low levels of driven current in the presence of fast ions, but that it can predicate much higher performance (both in normalised pressure, $\beta_{N}$, and energy confinement enhancement factor, $H_{98,y2}$) than forecast to achieve $Q=10$ in ITER whilst still avoiding NTMs.
Furthermore, sawtooth control can be achieved without strong sensitivity to the deposition position of the peak of the ECCD, provided it is inside $q=1$, making the real-time feedback control requirements less stringent than for direct NTM suppression.
Only a low level of ECCD power was required to avoid NTMs; in these ASDEX Upgrade plasmas, just 0.8MW of ECCD was sufficient to avoid NTM triggering up to $\beta_{N}=3.0$ with 14MW of auxiliary heating power.
The fact that a modest level of injected EC power could result in such a dramatic change in the sawtooth behaviour, despite the strong stabilising contribution of the energetic ions, suggests that the destabilising effect of increased local magnetic shear may be stronger than reference \cite{Porcelli1996} suggests; this is the case, for instance, in the stability criteria for the drift tearing mode in reference \cite{Cowley} where a fourth order dependence on $s_{1}$ appears.
Whilst these energetic particles represent up to approximately 20\% of the plasma pressure, this is still much less than expected in ITER, and definitive demonstration of the effectiveness of ECCD does require a larger fast ion fraction in future studies.

%---------------------------------------------------------------------------
\ack
The authors would like to acknowledge Dr V Bobkov for operation of the ICRH system. This work was partly funded by the RCUK Energy Programme under grant EP/I501045, the European Communities under the contract of Association between EURATOM and CCFE. The views and opinions expressed herein do not necessarily reflect those of the European Commission.

%---------------------------------------------------------------------------

%------------------------------------------------------------------------------
\pagebreak

%---------------------------------------------------------------------------------------
\end{document}